
\font\fourteenrm=cmr10 scaled \magstep2
\font\twelverm=cmr12 \font\ninerm=cmr9
\font\fourteenbf=cmbx10 scaled \magstep2
\font\twelvebf=cmbx12 \font\ninebf=cmbx9
\font\fourteenmi=cmmi10 scaled \magstep2
\font\twelvemi=cmmi12 
\font\ninemi=cmmi9
\font\fourteensy=cmsy10 scaled \magstep2
\font\twelvesy=cmsy10 scaled \magstep1
\font\ninesy=cmsy9
\font\fourteenex=cmex10 scaled \magstep2
\def\fourteenpoint{
 \textfont0=\fourteenrm \scriptfont0=\twelverm \scriptscriptfont0=\ninerm
 \textfont1=\fourteenmi \scriptfont1=\twelvemi \scriptscriptfont1=\ninemi
 \textfont2=\fourteensy \scriptfont2=\twelvesy \scriptscriptfont2=\ninesy
 \textfont3=\fourteenex \scriptfont3=\fourteenex
 \scriptscriptfont3=\fourteenex
 \textfont\bffam=\fourteenbf \scriptfont\bffam=\twelvebf
 \scriptscriptfont\bffam=\ninebf
  \def\rm{\fam0 \fourteenrm}%
  \def\bf{\fam\bffam\fourteenbf}%
  \normalbaselineskip=21pt
  \setbox\strutbox=\hbox{\vrule height14pt depth5.5pt width0pt}
  \normalbaselines\rm
}

\def\M{{\cal M}}
\def\L{{\cal L}}

\def\intinf{\int_0^\infty}
\def\intpi{\int^\pi_0}
\def\fpi{f_\pi^2}

\def\Fsin2{\sin^2 F(\rho)}
\def\sinF4{\sin^4 F(\rho)}

\mathchardef\lag="724C
\magnification=1200
\hsize=16.0truecm\hoffset=-0.3truecm
\baselineskip 10pt

\def\lc{\char'140}

\def\pl{{ Phys.\ Lett.\ }}
\def\np{{ Nucl.\ Phys.\ }}
\def\pr{{ Phys.\ Rev.\ }}
\def\prl{{ Phys.\ Rev.\ Lett.\ }}
\def\prp{{ Phys.\ Rep.\ }}
\def\cmp{{ Commun.\ Math.\ Phys.\ }}

\def\xt{{\hat {\bf x}} \cdot \vec \tau }

\vskip 4pc
\centerline{\bf { Phase structure of a quantized chiral soliton on S${}^3$}}
\vskip 4pc

\centerline{ Akizo Kobayashi  and Shoji Sawada$^\ast$}
\vskip 1pc
{\sl
\centerline{ Faculty of Education, Niigata University,
Niigata 950-21, Japan}
}
\vskip 1pc
{\sl
\centerline{$\ast$ Department of Physics, Nagoya University, Nagoya~464-01,
Japan }}
\vskip 6pc
\baselineskip 18pt
\parskip 6pt plus 1pt
\vskip 1pc
{\leftskip=2cm \rightskip=2cm
  A quantization of a breathing motion of a rotating chiral soliton on $S^3$
  is performed in terms of a family of trial functions for a profile function
of the hegdehog ansatz.
    We determine eigenenergies of the quantized $S^3$ skyrmion
    by solving the Schr\"odinger equation of the breathing mode for several
     lower spin and isospin  states varying the Skyrme term constants $e$.
   When $S^3$ radius is smaller than $2/ef_\pi$, where $f_\pi$ is the pion
    decay constant, we always obtain a conformal map solution as the lowest
     eigenenergy state.
   In the conformal map case, allowed states have only symmetric or
anti-symmetric wave function under inversion of a dynamical variable describing
the
breathing mode.  As the $S^3$ radius increases the energy splitting between
the symmetric and anti-symmetric states rapidly decreases and two states
become completely degenerate state.
    When the $S^3$ radius larger than $3/ef_\pi$, for the small Skyrme term
constant $e$ the lowest eigenenergy states are obtained with the
 profile function given by an arccosine form which is almost the
same to those of usual $R^3$ skyrmion.
  When the effects of the Skyrme term are weak, i.e. large $e$,
the lowest energy states are obtained by the  profile function of
 conformal map, which correspond to the \lc\lc frozen states" for the $R^3$
  skyrmion as the limit of $S^3$ radius $ \to \infty$.
 }
\vfil\eject

\baselineskip 19pt
\centerline{\bf \S 1. Introduction}

\par
\qquad
 A minimal extension of the nonlinear chiral Lagrangian can admit soliton
solutions called skyrmions which have been considered as baryons,
and the Skyrme model has been shown to provide a reasonable phenomenology
for static properties of baryons.${}^{1)}$
   Since Klebanov applied the skyrmion to the extended system of nuclear
matter with a finite baryon density, a variety of numerical studies of the
periodic arrays of the cubic lattice skyrmion in the flat space $R^3$ have
been made.${}^{2) \sim 5)}$
   These approaches, however, are technically so complicated and require
tedious numerical works due to the huge number of dense mesh points.
   The other approach is initiated by Manton where the flat space $R^3$ is
replaced by the 3-sphere.${}^{3),4)}$
  In this approach a single baryon on a 3-sphere $S^3(\rho)$ with radius
$\rho$ is just an approximation of baryonic matter with an average baryon
density of
$1/2\pi^2\rho^3$ and provides very good approximation of the
generic features including the phase transition of the $R^3$ lattice skyrmion
approaches.${}^{5)}$

     The most striking fact that is found from these studies is existence
of two  phases. In the $S^3$ skyrmion case, one is a phase where both baryon
number and energy densities are well localized around the north or south pole
of $S^3$ corresponding to the low baryon density matter.
 The other is a phase where these densities  are much more homogeneously
distributed and corresponds to the high baryon density matter.
    It is expected that transitions between the localized states and the
homogeneous states will be closely related to the breathing motion of skyrmion.
  So far, as we know, no studies of the $S^3$ skyrmion on quantum level
including
 breathing mode have been made.
   Recently, it has been shown by Yang and one of the author${}^{6)}$ that
for a certain region of the Skyrme term constant the generic features of the
$R^3$ skyrmion
 on quantum level is drastically changed from those on classical level
and that the rotational and breathing modes are completely frozen.
   Therefore, it is very interesting to study whether
quantum effects of the breathing motion may give rise to
 fundamental changes in energy levels and phase structures of the $S^3$
skyrmion.
   These generic features including the phase structure seem to show
close relations and striking similarities to the baryonic matter in the
flat space $R^3$ as known on classical level.${}^{5)}$

  In this paper we investigate the phase structure of a quantized chiral
soliton on $S^3$, where the effects of the quantization of breathing and
rotational modes are taken into account on a basis of a family of
trial functions for the profile function of the hegdehog ansatz.${}^{6),7)}$
  This approach is also a natural extension of the  analysis in Ref.~{6)}
for the $R^3$ skyrmion  to those for $S^3$ skyrmion system.

       This paper is organized as follows.
  In Sec. II the standard skyrmion on $S^3$ is quantized by taking account
of the breathing mode in addition to the spin-isospin rotation.
  In Sec. III on the basis of a family of trial functions for the profile
functions of the $S^3$ skyrmion we obtain explicit expressions of the
Lagrangian and the Schr\"odinger equation for the spin-isospin rotation
and the breathing motion.
  In Sec. IV we investigate the effects of quantization of the breathing mode
   by solving numerically the Schr\"odinger equation for the $I=J=0$ state of
    the $S^3$ skyrmion. In Sec. V we study the phase structure of the quantized
$S^3$ skyrmion obtained by solving the Schr\"odinger equation for spin-isospin
states $I=J=1/2,$ 3/2 and 5/2.
 The final section is devoted to discussion and conclusions.

   \vskip 4pc

\centerline{\bf \S 2. Breathing Motion of S$^3$ Skyrmion}

\par
\qquad
 In the standard Skyrme model${}^{3),4)}$ on $S^3$, the Lagrangian of a chiral
$SU(2)_L \times SU(2)_R$ nonlinear $\sigma$ model is given in the massless
pion limit as
$$
\eqalignno{
    \L  =& \L_A + \L_{\rm SK} ,  \cr
  \L_A =& {1\over 4}f^2_{\pi}{\rm Tr}(g^{\mu \nu} \partial_{\mu}
                    U(x) \partial_{\nu} U^{\dagger}(x)), &(1) \cr
    \L_{\rm SK} =& {1 \over 32e^2 }{\rm Tr}(g^{\mu \nu} g^{\kappa \lambda}
[U^{\dagger}(x)\partial _{\mu}U(x), U^{\dagger}(x)\partial_{\kappa} U(x)]
[U^{\dagger}(x)\partial _{\nu}U(x), U^{\dagger}(x)\partial_{\lambda} U(x)]),\cr
 }
$$
where $U(x)=\exp (2i\pi(x)/f_\pi)$ is the element of $SU(2)$ group,
$\pi(x)=\pi^a(x)\tau^a/2$ describes the pion field,
$f_\pi=93$MeV is the pion decay constant.
  The $\L_{\rm SK}$ term, called Skyrme term,${}^{1)}$ is introduced to
stabilize the soliton solutions where $e$ is a free parameter.
 Here $g^{\mu \nu}$ is the inverse of metric tensor $g_{\mu \nu}$ in a
curved space where $g_{0\mu}=\delta_{0\mu}$.
   Using the standard spherical coordinate $i=(\omega,\theta,\phi)$ on
$S^3(\rho)$,
the metric tensor $g_{ij}$ on the 3-sphere of radius $\rho$ are given as
$$
   g_{ij}=-\delta_{ij} g_i, \ \ \ g_\omega=\rho^2,\ \ \ g_\theta =
\rho^2 \sin^2\omega, \ \ \ g_\phi=\rho^2 \sin^2\omega\sin^2\theta.  \eqno(2)
$$
   The hedgehog ansatz for the classical static $S^3$ skyrmion${}^{1),3)}$ is
given by
 $$\eqalignno {
       U_0({\vec x})=& \exp [i\xt F(\omega)]   \cr
      {\bf \hat x} =& {\bf \hat e}_x \sin \theta \cos \phi +
{\bf \hat e}_y  \sin \theta \sin \phi + {\bf \hat e}_z \cos \theta , &(3) \cr
}
$$
where $F(0)=\pi$ and $F(\pi)=0$ for the baryon number $B=1$ sector.
 Substitution of $U_0$ into $U$ of (1) as
$$
        \eqalignno{
      & M_0= -\int \L (U_0 (\vec x)) \ dV,  \cr
      &dV=\sqrt{g}d^3x = \rho^3 \sin^2 \omega d\omega d\phi
                    \sin \theta d\theta, \ \ \ \ g={\rm det} g_{ij}, &(4) \cr
        }
$$
 leads to the static energy functional;${}^{3),4)}$
$$
\eqalignno{
    M_0[F]=& 2\pi \rho \int^\pi_0 d\omega \int^\pi_0 \sin \theta d\theta
 [\M_A + \M_{\rm SK}], \cr
      \M_A =& {f^2_\pi \over  2}(F^{'2} \sin^2 \omega +2\sin ^2F), &(5) \cr
      \M_{\rm SK} =& {1 \over 2 e^2 \rho^2} \big \{ {\sin ^4 F
               \over \sin^2 \omega} + 2(F' \sin F)^2 \big \}, \cr
}
$$
 where $F$ means $F(\omega)$ and $F'=dF(\omega)/d\omega$.
 The classical profile function $F_{cl}$ minimizes the functional
$M_0[F]$.

 The quantization of the rotational and breathing modes is performed by
the following replacement in the Lagrangian (1):${}^{8)}$
$$
        U({\vec x,t})=A(t){\exp}[i\xt F(\omega, \chi(t))]
                      A(t)^{\dagger},
         \eqno{(6)}
$$
 where $ A(t) \in SU(2)$ and $\chi(t)$ are the dynamical variables of
the spin-isospin rotation and the breathing motion respectively.
 It is convenient to express the breathing motion as $F(\omega,
\chi(t))=F(\mu)$  by use of a dimensionless variable $\mu=\tan
(\omega/2)/\chi(t)$ which assures invariant boundaries of the range of
variable $\mu$ for any $\chi(t)$ as $\mu=0$ at $\omega=0$ and $\mu=\infty$ at
$\omega=\pi$.
  If the boundary values of a given trial profile function $F(\mu)$ are
taken as $F(0)=\pi$ and $F(\infty)=0$, therefore, the breathing profile
function
$F(\omega,\chi(t))$ satisfies automatically the boundary conditions for
the baryon number $B=1$ sector on $S^3$, i.e. $F(\omega=0,\chi(t))=\pi$
and $F(\omega=\pi,\chi(t))=0$.
  It is worth noting that for the limit of
the 3-sphere radius $\rho \to \infty$,  $\mu$ is just equivalent to the
dimensionless
variable $r/R(t)$ of the flat spaces $R^3$, where $R(t)$ is regarded as a
dynamical variable for the breathing motion of the $R^3$ skyrmion.${}^{6)\sim
8)}$
  For the simultanious limit $\rho \to \infty$ and $\omega \to 0$, a radial
distance $r$
between a point $(\omega, \theta, \phi)$ on $S^3$ and the origin of $R^3$ is
approximated to $2\rho\tan (\omega/2)$ which is just the radial distance
between
the point $(\omega, \theta, \phi)$ and the north pole $( \omega=0 )$ of $S^3$.
   Thus, if we identify the dynamical variable $\chi(t)$ with $R(t)/(2\rho)$,
where $R(t)$ is a dynamical variable of the breathing motion in the flat space
$R^3$,${}^{8)}$
 The dimensionless variable $\mu=\tan (\omega/2)/\chi(t)$
behaves as
$$
     \lim_{\omega \to 0} \mu=\lim_{\omega \to 0}{ {\strut 2\rho \tan {
\displaystyle {\omega \over 2} } }
\over R(t)}={r \over R(t)}.  \eqno (7)
$$
  Now the profile function $F(\mu)=F(r/R(t))$ describes the breathing
motion of a localized skyrmion at the origin of $R^3$.
 These are the reason why we use the profile function $F(\mu)$ with
$\mu=\tan (\omega/2)/\chi(t)$.

 By substituting (6) to (1) we obtain the Lagrangian for the spin-isospin
rotation and the breathing motion of the chiral solitons given by
$$
   L= a(\chi){\dot \chi}^2 + \lambda(\chi) {\rm Tr}[{\dot A}
{\dot A}^{\dag} ] - V(\chi).
 \eqno (8)
$$
 Here $a(\chi)$ and $\lambda (\chi)$  are the $\chi(t)$ dependent
inertia of the breathing motion and the moment of inertia of the spin-isospin
rotation, respectively, and are given by
$$
\eqalignno{
   a(\chi)=&2\pi \int^\pi_0 d\omega \rho \big[ \fpi \rho^2 \sin^2
\omega  + {2 \over e^2} \sin^2F  \big] \big( {dF \over d\chi} \big)^2 , \cr
\lambda(\chi)=&{ 8\pi \fpi \over 3}\intpi d\omega \rho^3 \sin^2 \omega
 \big[ 1+{1 \over e^2  \rho^2}( F'^2+ {\sin^2F \over \sin^2\omega} ) \big ]
\sin^2F ,
&(9) \cr
}
$$
 where $F=F(\omega,\chi(t))$ and $F'=dF(\omega,\chi(t))/d\omega$.
It is noted that $dF/d\chi$ is rewritten as $-(\mu/\chi)dF(\mu)/d\mu $ by
use of an expression $F(\omega,\chi(t))=F(\mu)$ with $\mu=\tan
(\omega/2)/\chi(t)$.
  The $V(\chi)= M_0 [F]$ is the potential of the breathing motion obtained
by replacement of $F(\omega)$ by $F(\omega, \chi(t))$ in (5).

 According to the standard procedure of the quantization
based on the Lagrangian (8), after separation of the rotational mode,
we obtain the Schr\"odinger equation for the wave function of the breathing
mode, $\psi_J(\chi)$ for a $I=J$ spin-isospin eigenstate:${}^{6),8)}$
$$
[- {1 \over 4a(\chi)} \{ {d^2 \over d\chi^2 } + h(\chi) {d \over d\chi } \} +
V(\chi) +
{ J(J+1) \over 2\lambda(\chi)} -E_J] \psi_J (\chi) = 0, \eqno (10)
$$
where $E_J$ is an energy eigenvalue and
$$
h(\chi)={ 3\lambda'(\chi) \over 2\lambda(\chi) } - { a'(\chi) \over 2a(\chi) },
$$
and prime means differentiation with respect to $\chi$.

\vskip 4pc

\centerline{\bf \S 3.  Trial Profile Functions and
Quantized S${}^3$ Skyrmion}

\par
\qquad

  In order to obtain the quantized skyrmion solutions on $S^3$,
we have to solve the Schr\"odinger equation of the breathing motion (10) for
spin-isospin eigen states varying the profile function
$F(\mu)$ which gives  the most minimum eigenvalue $E_J$.
 To investigate for the wide range of variation of the profile
function we adopt the variational method in terms of a family of trial
functions $F(\mu;C)$ with a parameter $C\quad(C\ge 0)$, which is expressed
by an integral form as ${}^{7)}$
$$
   F(\mu; C)={ \pi \over N} \int_\mu^\infty { dx \over (x^2+1) \sqrt{x^2+C}},
   \eqno (11)
$$
   where
$$
   N=\intinf { dx \over (x^2+1) \sqrt{x^2 +C}}.
   $$
   The trial profile functions $F(\mu;C)$ satisfy the boundary condition
$F(0;C)=\pi$ corresponding to a soliton of the baryon number 1 and the
asymptotic condition for finite energy $F(\mu;C) \to 0$ for $\mu \to
\infty$. $F(\mu;C)$ with finite $C$ has the asymptotic
falloff $\propto 1/\mu^2$ for large $\mu$ which is required by the
Euler-Lagrange equation for the profile function in the chiral limit
regardless of the stabilizer. This asymptotic falloff also agrees
with that of the static meson theory in the massless pion limit.

    In the usual $R^3$ skyrmion the function with $C=1$, $F(\mu;C=1)$, is
just the same as the one derived by Atiyah and Manton${}^{9)}$ from the
instanton solution and it is the nearest trial function
to the classical numerical solution $F_{cl}(\mu)$ among the family $F(\mu;C)$.
    It gives a static energy  only 0.9\% above that of numerical
solution.${}^{6)}$
 The case with $C=2$ is a trial function of arccosine form introduced by
Igarashi, Otsu and two of the present authors,${}^{10)}$ which gives also
a very good approximation of the exact static profile function $F_{cl}(\mu)$
and gives a static energy only 1.0\% above that of the numerical solution.

By use of the trial profile function (11) we obtain the inertias $A(\chi)$
and $\lambda(\chi)$ from (9) and the potential $V(\chi)=M_0 [F]$ from (5).
These inertias and potential are dependent of the parameter $C$. Varying
this parameter $C$ we solve the Schr\"odinger equations (10) for the
breathing motion with definite spin and isospin $I=J$.
\par
In the limit of the $S^3$ radius $\rho \to \infty$ the $S^3$ skyrmion
tends to a usual $R^3$ skyrmion. In this case, when the Skyrme term
contribution is small
(i.e. when the Skyrme term constant $e >5.0$ ), there appear the `` frozen
 states" in which all the freedoms of the spin-isospin rotation and the
breathing motion are completely frozen.
    This is because  the state with  infinite inertias of the
breathing motion and with infinite moment of inertia of spin-isospin rotation,
so that the contributions from the kinetic energies of both modes become zero
and the $R^3$ skyrmion energy comes only from the potential energy of the
breating
mode.  The profile function $F(\mu)$ in the frozen state is given by
$F(\mu;C=\infty)$ which gives the  most minimum eigenenergy $E_{\infty}$ in the
region  $e > 5.0$. In this region of $e$ the classical profile function
 $F_{cl}(\mu)$ or the trial function
$F(\mu;C=1)$ gives local maximum eigenenergies.
\par
   On the other hand when the Skyrme term constant $e$ is small the frozen
state does not appear for the small spin-isospin states and the classical
profile function which is approximated by $F(\mu;C=1)$ gives most minimum
eigenenergies.
\par
  We confirmed numerically that in the case of
$S^3$ skyrmion a similar situation occurs as in the $R^3$ skyrmion.
    The eigenenergies of $S^3$ skyrmion with the definite spin-isospin $I=J$
become minimum at $C=1$ or at $C=\infty$ depending on the parameter $e$.
  In the following, however, we calculate numerically for two special values of
parameter $C$,
i.e. $C=2$ and $C=\infty$. The value $C=2$ is a substitute of $C=1$.
   The reason for solving with $C=2$ instead of $C=1$ lies on the fact that
we can  obtain analytically integrated results of the potential $V(\chi)$
and inertias $a(\chi)$ and $\lambda(\chi)$ for $C=2$ as well as $C=\infty$
but not for $C=1$.
 The difference between the eigenenergies for $C=2$ and $C=1$ is expected
within 1\% as in the case of the $R^3$ skyrmion.

\noindent
  {(i){\sl  Breathing motion, inertias and potential for C = $\infty$} }

  The tial profile function for $C=\infty$
$$
F(\mu;C=\infty)=\pi \big( 1 - { 2 \over \pi } \arctan \mu \big)
    ,\eqno (12)
$$
 is rewritten as
$$
{{\strut \tan \displaystyle {\omega \over 2} } \over \chi(t)}=\cot
{F(\mu;C=\infty)\over 2}=\mu.
    \eqno (12')
$$
  which is known as a conformal map.${}^{3),4)}$

  Now we investigate the breathing motions of the profile function and baryon
density given by the conformal map:
$$
  F(\omega,\chi(t)) = 2 \,  {\rm arccot}\, \mu.
  \eqno (12'')
$$
  In Fig.~1(a) we show the breathing profile functions for various values of
$\chi(t)$, i.e. $\chi=1/5,\, 1/3, \, 1, \, 3$ and 5.
  The solid line represents the conformal map profile functions.
 The diagonal line represents the identity map $F(\omega,1)=\omega$.
 It is noted that the peak of the profiles $F(\omega,\chi(t))$ at $\omega=0$
  shrinks or  spreads out as the values $\chi(t)$ decreasing or increasing
  from 1.

   The baryon density for the $\chi(t)$ at the radial angle $\omega$ on $S^3$
is defined as
$$
    2\pi^2 j^0 (\omega) \rho^3 = {\sin^2 F(\omega, \chi(t)) \over
     \sin^2 \omega } {d F(\omega,\chi(t)) \over d \omega}
  \eqno (13)
$$
where $j^0 (\omega) $ is the baryon number density.
 The solid line in Fig.~1(b) represents the baryon number density given by
  the conformal map for various $\chi(t)$ values.
   The distributions of the baryon number density for the conformal map
profile function have a exact symmetry.
 It is easy to prove by using the explicit form for $F(\omega,\chi(t))$
of $(12'')$ that the effect of an inversion $\chi \to 1/\chi$ in (13) is
equal to the one of a transformation $\omega \to \pi-\omega$ by which
the north semisphere of the $S^3$ is transformed on to the south semisphere.
  Thus, the peaked shapes of the baryon density around the north pole
for $\chi=1/2$ and 1/3 are exactly mirror symmetric to those of the
corresponding one around the south pole for $\chi= 2$ and 3, respectively.
  The identity map $F(\omega,1)=\omega$ gives completely flat baryon density
over the whole 3-sphere as shown by the solid line of $\chi=1$ in Fig.~1(b),
i.e. $2\pi^2 j^0 (\omega) \rho^3=1$.
   Thus, the breathing motion expressed by the dynamical variable $\chi(t)$
means that distributions
of the baryon density flip-flops between a higher peak at the north pole
(e.g. $\chi=1/2$) through completely flat one ($\chi=1$) and to higher one
at the south pole ($\chi=2$), and vice versa, as shown in Fig.~1(b).

  Here, by substituting (12) to (9), we obtain integrated expressions of the
 inertias $a(\chi)$ and  $\lambda (\chi)$ and the potential
$V(\chi)$ as follows,
$$
   \eqalignno{
     a(\chi)=&{4\pi^2 \over f_\pi e^3}{ \Lambda (\chi) \over \chi^2}, \cr
   \lambda(\chi)=&{16\pi^2  \over 3f_\pi e^3} \Lambda (\chi), \cr
   V(\chi)=& {6\pi^2f_\pi \over e} \big( {2L \over {\strut \chi +
\displaystyle {1 \over  \chi} +2 } } +
      {1 \over 4L}(\chi + {1 \over \chi}) \big),         &(14)  \cr
 \Lambda (\chi)=&{1 \over {\strut (\chi+ \displaystyle {1 \over \chi} +2 )^2 }
}\big(3L^3+ L(\chi+{1 \over \chi}+4) \big).
}
$$
where $L=ef_\pi \rho$ is the dimensionless $S^3$ radius and will be use instead
 of the radius $\rho$ of the 3-sphere $S^3$ in the following.

  We notice that under an inversion;
$$
\chi \to {1 \over \chi},
    \eqno (15)
$$
there exists a symmetry as follows;
$$
   \eqalignno{
 V(\chi)=&V \big( {1 \over \chi} \big), \cr
\lambda(\chi)=&\lambda \big( {1 \over \chi} \big),  &(16)  \cr
 a(\chi)\big({d \chi \over dt}\big)^2=& a({1 \over \chi})\big({d \over dt}
 {1\over \chi} \big)^2 .
}
$$

    Therefore for the conformal map profile function (12), the Lagrangian (8)
and
the Schr\"odinger equation (10) are invariant under the inversion (15).
Hereafter, we call this invariance as the inversion symmetry.
 In Fig.~2(a), the potential $V(\chi)$ of the conformal map case (12) is shown
for  varying the radius $L$. The potential exhibits  inversion symmetry.
 In the region $L \le {\sqrt 2}$ the potential $V(\chi)$ has single valley
 symmetric for the variable  ln$\chi$ and the potential $V(\chi)$  takes the
absolute minimum value at $\chi=1$ for fixed value of $L$.
  This conformal map with $\chi=1$ called identity map  realizes a
homogeneous distribution of the baryon number density over the whole 3-sphere
as shown in Fig.~1(b).
  In these $L$ region, the  minimum of potential for the conformal map profile
function is given by
$$
    V_{\rm min}({\rm conformal})={3\pi^2 f_\pi \over e}(L+{1 \over L}),
\ \ L \le \sqrt{2}, \eqno (17)
$$
  It is noted that this really gives the energy of exact solution,
i.e. the static classical energy $E_{cl}$, as shown in Ref.~3 and 4.
  Here, it is also worth noting that the theoretical lower bound energy
$ 6\pi^2 f_\pi/e$ is realized at $\chi=1$ and $L=1$,${}^{3),4)}$
i.e. the absolute minimum of the potential valley in Fig.~2(a).

In the region $L \ge {\sqrt 2}$, the potential $V(\chi)$
behaves as the double well potential of the variable ln$\chi$ for a fixed
$L$ and has local maxima at $\chi=1$  and
two local minima  at $\chi=\chi_{lm}=\sqrt 2 L-1-\sqrt {2(L^2-\sqrt 2 L)} \le
1$ and at $\chi=1/\chi_{lm}=\sqrt 2 L-1+\sqrt {2(L^2-\sqrt 2 L)} \ge 1$.
 The minimum value of potential energy is given as a function of $L$
by${}^{3),4)}$
$$
     V_{\rm min}({\rm conformal})={3\pi^2 f_\pi \over e}(2 \sqrt{2}-{1 \over
L}),
\ \ L \ge \sqrt{2}.   \eqno (18)
$$
This minimum value is considerably larger than those of the exact
solution${}^{3),4)}$,
which is approximated well by the solution for the arccosine profile
function with $C=2$ given in the following.

\noindent
{(ii){\sl Breathing motion, inertias and potential for C = 2} }

   The trial profile function for $C=2$ in (11) is  written by  arccosine form;
$$
      F(\mu;C=2)= \arccos \big( 1-{2 \over (1+\mu^2)^2} \big). \eqno (19)
$$

 The dotted line in  Fig.~1(a) show the breathing profile functions
of the arccosine form for  values of $\chi(t)=1/5, \, 1/3, \,  1, \, 3$ and 5.
  The dotted lines of Fig.~1(b) display the corresponding baryon number
  densities.
The distributions of the baryon number density has no symmetry under the
transformation $\chi \to 1/\chi$ and $\omega \to \pi-\omega$
in contrast to those for the conformal map profile function.
  The baryon density has a peak at the north pole for $\chi=$1/3, 1/2
and 1, but has a bump and dip around the south pole for $\chi= 3$ and 2
$( \ge 1.6)$ as shown by dashed line in Fig.~1(b).

  Now, we obtain analytically integrated expressions by substituting (19) to
(9) for the inertias $a(\chi)$ and $\lambda (\chi)$ and potential $V(\chi)$ as
follows:
$$
\eqalignno{
 a(\chi)=&{2\pi^2 \over f_\pi e^3} \big( L^3 A_1(\chi)+L B_1(\chi) \big), \cr
 \lambda(\chi)=&{8\pi^2  \over 3f_\pi e^3}  \big( L^3 A_2(\chi)+L B_2(\chi)
\big),  &(20)  \cr
 V(\chi)=&{2\pi^2f_\pi \over e} \big( L A_3(\chi)+{B_3(\chi) \over L} \big) ,
\cr
 }
 $$
 where
 $$
 \eqalignno{
 A_1(\chi)=&{8\chi \big( 32-20{\sqrt 2}+(8+{\sqrt 2})\chi+6\chi^2 \big)
\over {\sqrt 2}(1+\chi)^4(1+{\sqrt 2}\chi)^3 }, \cr
 B_1(\chi)=&{ 3+18\chi+42\chi^2+42\chi^3+7\chi^4 \over 2\chi(1+\chi)^6 }, \cr
 A_2(\chi)=&{ \chi^3(7+12\chi) \over (1+\chi)^6 },   &(21)  \cr
 B_2(\chi)=&{
\chi(157+1256\chi+4251\chi^2+7632\chi^3+7243\chi^4+3176\chi^5+397\chi^6)
\over 128(1+\chi)^8}, \cr
 A_3(\chi)=&{8\chi \big( 3-2{\sqrt 2}+(2-{\sqrt 2})\chi \big) \over
(1+\chi)^2(1+{\sqrt 2}\chi) } +  { \chi(3+12\chi+11\chi^2) \over 2(1+\chi)^4 },
\cr
 B_3(\chi)=&{ {\strut 23 \big(11\chi+ \displaystyle {27 \over \chi}  \big) }
 \over 512 }.\cr
}
$$
   In this arccosine profile function case there exists no special symmetry in
contrast with the
 conformal map case.
  As shown in Fig.~2(b) ,
  in the region $L \le {\sqrt 2}$ the potential $V(\chi)$ has also a single
well
but  with no special symmetries and  has a minimum at $\chi \sim 1.6$.
   The minimum of the potential, however, is considerably larger than that of
 the conformal solution given by (17) in this small $L$ region.
    This means that in classical level, the stable soliton
solution for $L \le {\sqrt 2}$ is given by the profile function of the
identity map (the conformal map with $\chi=1$).

In the region $L \ge {\sqrt 2}$, the $V(\chi)$ behaves as a double well
potential which  has a local maximum at $\chi \sim 1.6$ (saddle point)
and two local minima  with two different depths.
  The smaller one of these two minima exists in the region of $\chi < 1.6$.
This minimum value given by the arccosine profile function (19) gives
a very good approximation of the exact solution within about 1\% deviation.
  Thus, for $L \ge {\sqrt 2}$ the exact classical solution given
by the $F_{cl}$  is  approximated well by the arccosine profile function
$F(\mu;C=2)$.

\vfill
\eject

\centerline{\bf \S 4. Effects of Quantization of Breathing mode}

\par
\qquad
\noindent
 In order to examine the effects of quantization of the breathing mode we
 study here the solutions of the Schr\"odinger equation for the breathing
 motion of the $I=J=0$ state.

 In Fig.~3 we display the arbitrarily normalized wave functions
$\psi_{J=0}(\chi)$ in a region $ 0 \le \chi \le 1$ as a typical example of
the solutions of conformal map with $C=\infty$.
   The solid line and the dashed line represent the symmetric state and
anti-symmetric state, respectively, for $L=1.3$, at which
the eigenenergy of the symmetric state takes the lowest value among all $L$
region. The value is  $E_{I=J=0} = 870$Mev for $e=7$. The potentials
  $V(\chi)$ for $L=1.3$ and $L=10$ with $e=7$ also shown by dotted lines in
Fig.~3.
The potentials and the wave functions in the region $\chi \ge 1$ are given
from those in $0 \le \chi \le 1$ by $V(1/\chi)$ and $\pm \psi_{J=0}(1/\chi)$,
respectively.    For $L = 10$ we get  degenerate states as shown by dot-dashed
line
in Fig.~3.
  The absolute values of wave functions have sharp peaks at $\chi=\chi_{lm}
\sim 0.03$ and $\chi=1/\chi_{lm} \sim 33$ where the potential $V(\chi)$ takes
the minimum value
as shown in Fig.~3 for $0 \le \chi \le 1$.
   The probability distributions to find the $\chi$ is so localized at
$\chi \simeq \chi_{lm} $ and $\chi \simeq 1/\chi_{lm} $, therefore, the
quantum tunneling effect between these two regions is considered to be
negligible.

    In Fig.~4, we show calculated results of eigenenergy $E_{I=J=0}$ as a
function of $L$ for several values of the Skyrme term constant $e$.
      Solid and dotted lines represent those of the symmetric and
anti-symmetric
solutions, respectively, for $C=\infty$ (conformal map).
    Two lines of eigenenergies are completely degenerate for large $L$-region.
  Dot-dashed lines represent eigenvalues $E_{J=I=0}$ for $C=2$ (arccosine
form).
  For $e=7$, we display
  also in Fig.~3 by  dashed lines the minimum values of the potential
$ V_{\rm min} ({\rm conformal})$ given by (17) and (18) for the
conformal map profile function with $e=7$.
 Therefore, the difference between the solid line and the dashed line
represents  contribution from the kinetic energy of  breathing motion.

   From these results for $I = J = 0$ the effects of quantization of the
breathing mode are summarized as the followings:

\noindent
  (i) In classical level, there exist two phases.${}^{3)}$
     One is the phase where both baryon number and energy densities are well
      localized at the north or south pole of $S^3$ in the region
        $L >\sqrt{2}$, which corresponds to the $R^3 $ skyrmion with usual
         profile function numerically solved.
 The other is the phase where these densities are homogeneously distributed
  in the the region $L<\sqrt{2}$. In this case the profile function is given
  by that of the identity map.
 In the quantum level, the quantum breathing motion in terms of the profile
 function for conformal map on $S^3$ means quantum flip-flop between a
 localized distribution at north pole ($\chi<1$) and a localized one at
  the south pole ($\chi>1$) through a uniform distribution (identity map with
   $\chi=1$).

\noindent
(ii)   As shown in the previous section and also in Fig.~2(a), the quantum
breathing motion  is described by a single-well potential in the region
$L \le \sqrt 2$, but it is described by a double-well potential in the region
$L \ge \sqrt 2$.
  In the conformal map case the height of the potential barrier at $\chi=1$
  from the bottom of potential increases from 0 to $\infty$ as $L$
   increases from $\sqrt 2$  to $\infty$.
 Thus, the ground state wave function changes smoothly from the one which has
 a peak at $\chi=1$ to the one with double peaks at around potential minimum
 points $\chi_{lm}$ and  $1/\chi_{lm}$  as $L$ increasing from 0 to $\infty$.

\noindent
  (iii) The one of new features of quantization of the breathing mode is
   the appearance of pairs of
eigenstates; symmetric and anti-symmetric under the inversion (15)
 for the conformal map profile function.
    In the region of small $L$ less than $\sim 2$ the splitting between these
two energy levels is rather large due to the lack of energy barrier between
small and large $\chi$ regions.
   On the other hand, in the region of large $L$ larger than $\sim 3$
this pair of states has an almost degenerate energy levels.
  This means the potential barrier around $\chi = 1$ become high enogh
to forbid the penetration between the small and large regions of $\chi$.
   As for the arccosine case, the potential $V(\chi)$ shows a
single-well or double-well  behaviors depending on the sphere size $L$
but does not show any symmetry such as given by (15).
  Therefore, the eigenenergies for the arccosine case have no degeneracy.
In Fig.~4 only lower energy solutions are shown.

\noindent
  (iv)  For small Skyrme term constant $e$ and in the large $L$ region
  the arccosine profile function gives lower energy than the conformal map
case.
For $e=3$, 4 and 5 the arccosine case gives lower eigenenergy
     solution in a region $ L > 3$.
   For $e=7$ the energy of the arccosine case is slightly lower than the one of
the conformal map case in a region $3 < L < 20$, but the conformal map case
gives the lowest energy solution for $L > 20$ .
       For the limit $L \to \infty$ we get the same results as Ref.~6 on
$R^3$, i.e. for values of $e$ larger than a critical value
$e_0=5.6$ all minimum energy values of spin-isospin $J$ states, $E_J$,
 degenerate to a value
lim$L \to \infty E_{C=\infty}(e)=$ $7.788/e$ GeV of the conformal map case.

\noindent
 (v) As for the classical $S^3$ skyrmion, the  lowest energy is
always realized by the radius $L=1$ for any $e$.
  For the quantized $S^3$ skyrmion, however, the radius $L$ giving the
lowest energy  $E_0^{min}$ increases slightly  as $e$ increase.
  That is, the lowest energy is $E_0^{min}=1.902$GeV at $L=1.1$ for $e=3$,
    $E_0^{min}=1.456$GeV  at $L=1.1$ for $e=4$,  $E_0^{min}=1.185$GeV
at $L=1.2$ for $e=5$ and $E_0^{min}=0.870$GeV at $L=1.3$ for $e=7$,
respectively.

\vskip 4pc
\centerline{\bf{ \S 5.  Phase Structure of $S^3$ Skyrmion}}
\par
\qquad
  Now we discuss  the quantized $S^3$ skyrmion obtained by solving the
Schr\"odinger equation with $I=J= 1/2$, 3/2 and 5/2.
In Fig.~5(a) $\sim$ (d) we show the $L$-dependence of eigenvalues $E_J$
obtained for $J=1/2,$ 3/2 and 5/2 and for the Skyrme term constant $e=$3, 4, 5
and 7.
These eigenenergies of the symmetric and anti-symmetric states for conformal
 map profile function are represented respectively by the solid and dotted
linesand
 almost degenerate in the large $L$ region. The dot-dashed lines in Fig.~5
represent the lowest eigenenergies for the arccosine profile function.

  For $e=3$, as shown in Fig.~5(a), the  arccosine profile function realizes
the lowest energy solutions for all spin-isospin states $I=J=$1/2, 3/2 and 5/2
in the region
  $L \ge 2$.
   In the region $L \le 2$ the lowest energies are
obtained by the conformal map profile function for all spin-isospin states.
Then a  phase transition occurs at $L \simeq 2$.

 Fig.~5(b) exhibits the $L$-dependence of eigeneneries for $e=$4. In this case,
   for spin-isospin states $I=J=1/2$  and 3/2,
 there occurs  a phase transition at $L \simeq 2.5$ i.e.
   the conformal and  the arccosine profile functions give lower energies
    in the region $L < 2.5$ and  $L > 2.5$, respectively.
On the other hand, for $I=J=5/2$  the conformal map profile function always
gives the lowest energy.

  For $e=5$, as shown in Fig.~5(c), in a  region $L > 2.5$,  the arccosine
profile
function  gives the lowest energy solution for $I=J=$1/2 but  the conformal
map profile function gives the lowest one for $I=1/2$ in the region $ L< 2.5$
  and for $I=J=3/2$ and 5/2 in all region of $L >0$.
  Then the phase transition occurs at $L \simeq 2.5$ only for $J=1/2 $, but
  not for $J=3/2$ and 5/2.

  Fig.~5(d) shows for the case $e=7$.
   For all spin-isospin states with $J=1/2, 3/2$ and 5/2,  the conformal
map profile function  always gives the lowest
energy solutions in whole $L$ region.

We have considered hitherto the case where the radius of $S^3$ is fixed $a$
$priori$ by some condition given from outside. Now we turn to a rather academic
or fictitious problem where the isolated
$S^3$ skyrmion is putting on the vacuum and its radius $\rho$ or $L$ is
determined by itself. In this case we have to regard the sphere radius $L$ as
a dynamical variable and to find minimum eigenenergies by solving Shr\"odinger
equation for both $L$ and $\chi$. If we assume that the radius $L$ is an
 adiabatically variable of the $S^3$ skyrmion, then we have to search the
  lowest eigenenergies varying the radius $L$.
    In Table I we list the local minimum energies $E_J$ in the small $L$ region
of
symmetric states and the values of $L$ which give the minimum energies for
$e=3,$ 4, 5 and 7.
Fig.~6 shows $e$-dependence of these local minimum eigenenergies $E_J$ for
$J=0$, 1/2, 3/2 and 5/2.
  As shown by solid lines of Fig.~5(a) $\sim$ 5(d), in the limit of $L=\infty$
   the eigenenergies  of the any spin-isospin states with conformal profile
    function converge to the same energy $E_\infty=$ $7.788/e$ GeV which is
the energy of the frozen state of the $R^3$ skyrmion.
    The classical $S^3$ skyrmion energy $E_{cl}=6 \pi^2 f_{\pi}/e$ and
     $E_{\infty}$ are also shown by dashed and dot-dashed lines.
     If the $S^3$ skyrmion with definite spin-isospin which is represented by
     the solid lines in Fig.~6 above the dot-dashed line of $E_{\infty}$,
the radius of $S^3$ skyrmion increase infinitly and the skyrmion given by the
conformal  profile functions disperses. The energy of $I=J=1/2$ state is always
smaller than $E_{\infty}$ in the region $e$. For $e>5$ the $S^3$ skyrmion with
 $I=J=3/2$ disperses while for $e<5$ the $I=J=3/2$ skyrmion given by the
conformal
 map profile function has most minimum energy. The $I=J=5/2$ $S^3$ skyrmion
 of conformal map becomes most minimum energy state for $e < 4.1$.
      Occurrence of infinite dispersion of the quantized $S^3$ skyrmion
is recognized by the $e$ and $L$ dependence of the kinetic energies of the
rotational and breathing modes and the potential of breathing mode. As given
in (14) and (20) both inertias $a(\chi)$ and $\lambda(\chi)$ are proportional
to  $e^{-3}$ and increase as  $L$ increases. Then the kinetic energies of the
 rotational and  breathing modes to $e^3$ and decrease as the $L$ increases.
    On the other hand the potential energy of the breathing mode is
proportional to $e^{-1}$.
    Therefore in the large $e$ region where the contributions of
 kinetic energies dominate, increase of $L$ reduces total eigenenergy and
eventually the $S^3$
    skyrmion disperses as $L \to \infty$.

As listed in Table I, the values of minimum points $L$ which give the lowest
 eigenenergies $E_J$ are shifted toward larger values from that of the
classical
 one, $L=1$. This is also explained by the quantum effects of the
kinetic energies of spin-isospin rotation and breathing motion.
   \vskip 4pc

\centerline{\bf \S 6. Discussion and Conclusions}

\par
\qquad
   In this paper we have investigated the quantum effects associated with
rotation and breathing motions of a $S^3$ skyrmion making use of a family
of trial profile function $F(\mu;C)$ with $\mu = \tan (\omega/2)/ \chi(t)$
where $\chi(t)$ represents the dynamical variable of breathing motion of
skyrmion.

  In the classical level, for $L \le \sqrt 2$,
  we have a $S^3$ skyrmion
 whose profile function is given by identity map $ F(\omega)=\omega$ and
its  energy density is distributed homogeneously inside $S^3$.
In this case the baryon number density is $1/2\pi^2 \rho^3=e^3 f_{\pi}^3/2\pi^2
L^3 >
e^3 f_{\pi}^3/4 \sqrt{2}\pi^2 $. If the Skyrme term constant $e$ is identified
to
the coupling constant $g_{\rho \pi \pi }$=5.8, then the baryon number density
is
 larger than 0.37fm${}^{-3}$ which  corresponds to high baryon number density
matter
 and considered as deconfined state.${}^{3),4)}$
  As shown in the preceding sections,
 in the region $L \le \sqrt {2}$ where
the potential $V(\chi)$ of the breathing motion has single valley at $\chi=1$,
 the quantiezed  $S^3$ skyrmion is realized by the conformal map profile
function.
 The ground state wave function of breathing motion of the  $I=J=0$ $S^3$
skyrmion is
   shown by a full line in Fig.~3 which is symmetric under inverse
transformation (15).
  As in the classical level  this state is considered as the deconfined phase
of baryonic matter,
then the breathing motion obtained here can be regarded as a part of the
breathing motion
 of the baryonic matter.

In the region $L \ge \sqrt 2$,
     for most region of parameters $e$ and $L$
     and for the states with lower spin-isospin,
 we obtain  a localized skyrmion in the classical level whose profile function
is given by
 arccosine form (19) and considered as the confined state.${}^{3),4)}$
     The arccosine profile function is an approximation of the Atiyah-Manton
profile function given by (11) with $C=1$.
 In this case the potential $V(\chi)$ has two valleys as shown in Fig.~2(b) and
the one
at  $\chi$ smaller than 1.6 is deeper than the one at $\chi$
 larger than 1.6.  Then the wave function of the lowest energy is localized at
 very  small $\chi$ and the quantized $S^3$ skyrmion is more sharply confined
at
 the north pole of $S^3$.
  In the quantum level, however, there occurs for a certain region of the
parameters
$e$ and $L$ and for the spin-isospin states  that the  conformal
   profile function gives  the lowest energy state of $S^3$ skyrmion.
  We take $L=10$ as an example. The $I=J=1/2$ $S^3$ skyrmion is given by the
 arccosine profile function in the region $e < 6.5$. Only in the region $e>
6.5$ the
 conformal profile function gives stable $S^3$ skyrmion. The $S^3$ skyrmion
  with $I=J=3/2$ is realized by the arccosine and the conformal profile
functions
in the regions $e<4.8$ and $e>4.9$, respectively. For $I=J=5/2$ $S^3$ skyrmion
state
 the conformal profile function gives the lowest energy in the region $e>
  4$ and the arccosine profile function does in the region $e<4$. In general
 in the region of small $e$ corresponding to strong stabilizer, the lower
spin-isospin states are given by the arccosine (or Atiyah-Manton) form of
  profile function and localized at the north pole of $S^3$ for large $L$.
When the Skyrme term constant $e$ increases, i.e. the contribution of
stabilizer
 becomes weak, the $S^3$ skyrmions change to the ones given by the conformal
map corresponding to
  deconfined system from higher spin-isospin states and eventually all
spin-isospin states.

  The conformal map profile function corresponds to the `` frozen states" on
  the $R^3$ skyrmion where the inertias $a(\chi)$ and   $\lambda(\chi)$ of the
   breathing and rotational modes are infinite.${}^{6),11)}$  However, since in
   the case of the $S^3$  skyrmion the value $L$ is finite then the inertias
$a(\chi)$ and
 $\lambda(\chi)$  are also finite contraly in the case of $R^3$
    skyrmion, the `` frozen states" do not appear any more.

In conclusion, there appears two types of breathing motions, one is the
breathing motion
of the profile function given by conformal map, and the other is
those  approximated by the arccosine profile function.
    For strong stabilizer the lowest eigenenergy state
in large $L$ region  is given by the breathing motion of
arccosine profile function.
  For the weak stabilizer the eigen state of (10) for the conformal map profile
function
 always gives the lowest energy in all $S^3$ radius $L$, which
continuously connects to the \lc\lc frozen states"${}^{6),11)}$ on $R^3$
in the limit $L \to \infty$.
  Howeve, in the small $L$ region, the lowest eigen state is
always given by the breathing motion of the conformal map profile function.

  Since the $S^3$ skyrmion approaches provide very good modeling of the
periodic arrays of the cubic lattice skyrmion in flat space $R^3$ as shown
by various studies in the classical level,${}^{5)}$ above results are not
specific ones of the  $S^3$ skyrmion, but also essential features of the
quantized lattice skyrmions in flat space $R^3$.
  For example, the quantum breathing motions on $S^3$ around the uniform
distribution (identity map with $\chi=1$), correspond to the quantum
fluctuations
of the baryon density around the homogeneous distribution with the
half-skyrmion
 symmetry${}^{5)}$ of the periodic array of the cubic lattice skyrmion in the
flat space $R^3$.
   Thus, it may be possible to find these remarkable features in the dense
nuclear matter.
   \vskip 3pc

\centerline{\bf Acknowledgements}
\par
\qquad
 We would like to thank H. Otsu for helpful discussions and comments.

\vskip 3pc

\centerline{\bf References}

\vskip 1pc

\item{1)} T. H. R. Skyrme, Proc. Roy. Soc. {\bf A260},127 (1961);
\np {\bf 31},556 (1962).\hfill\break
G. S. Adkins, C. R. Nappi and E. Witten,\np {\bf B228},552 (1983).\hfill\break
I. Zahed and G. Brown, \prp {\bf 142},1 (1986).

\item{2)} I. Klebanov, \np {\bf B262},133 (1985).\hfill\break
E. W${\rm \ddot u}$st, G.E. Brown and A.D. Jackson, \np {\bf A468},450 (1987).

\item{3)} N. S. Manton, \cmp {\bf 111},469 (1987).

\item{4)} H. Reinhardt and B. V. Dang, \pr {\bf D38},2881 (1988).

\item{5)} M. Kugler and S. Shtrikman, \pl {\bf B208},491 (1988); \pr {\bf D40},
3421 (1989).\hfill\break
 A. D. Jackson and J. J. M. Verbaarschot, \np {\bf A484}, 419
(1988).\hfill\break
L. Castillejo, P. S. J. Jones, A. D. Jackson, J. J. M. Verbaarschot and A.
Jackson, \np {\bf A501}, 801 (1989).\hfill\break
A. D. Jackson, C. Weiss, A. Wirzba and A. Lande, \np {\bf A494}, 523 (1989).
\hfill\break
 T. S. Walhout, \np {\bf A484} (1988) 397, \pl {\bf B227}, 10 (1989).

\item{6)} S. Sawada and K. Yang, \pr {\bf D44},1578 (1991).

\item{7)} A. Kobayashi, H. Otsu, and S. Sawada, \pr {\bf D42},1868 (1990).

\item{8)} P. Jain, J. Schechter, and R. Sorkin, \pr {\bf D39},998(1989).

\item{9)} M. F. Atiyah and N. S. Manton, \pl {\bf B222},438 (1989).

\item{10)} Y. Igarashi, A. Kobayashi, H. Otsu and S. Sawada, \pl{\bf B195}, 479
(1987).

\item{11)} A. Stern, \prl {\bf 59}, 1506 (1987).

\vskip 3pc

\centerline{\bf Figure Captions}
\item{ FIG.~1}  (a) The breathing profile functions for $\chi=$1/5, 1/3, 1, 3
and  5. The solid line represents the conformal map profile functions and
the dashed line represents the arccosine profile function.
(b) The baryon density as a function of the radial angle $\omega$ for
$\chi=$1/3, 1/2. 1, 2 and 3. The solid line (dashed line)
represents the baryon density for the conformal map profile function
(the arccosine profile function).
\item{ FIG.~2} The potential $V(\chi,L)$ as a function of two variable $\chi$
and $L$; (a) the potential for the conformal map profile functions.
(b) the potential for the arccosine profile function.
\item{ FIG.~3}  The potentials $V(\chi)$ in GeV and the arbitrarily
normalized wave functions $\psi_0 (\chi)$ of the spin-isospin state with
$I=J=0$ for the Skyrme term constant $e=7$ plotted against the variable $\chi$.
 The dotted line shows the $\chi$ dependence of the potential $V(\chi)$ in GeV
for  $L=1.3$ or for $L=10$.
  The solid line (dashed line) represents symmetric (anti-symmetric)
wavefunctions $\psi_0 (\chi)$ for $L=1.3$, and dot-dashed line represents a
wavefunction for the almost degenerate state for $L=10$.
\item{ FIG.~4}  The eigenenergy $E_{I=J=0}$ in GeV of three types of eigen
states as a functions of $L$ for various Skyrme term constants $e$;
the symmetric states (solid line), anti-symmetric states (doted line) and
almost
 degenerate states (solid line) for conformal map profile function, and almost
degenerate states (dot-dashed line) for the arccosine profile functions.
The dashed line represents the minimum value of the potential as a function
of $L$ for the conformal map profile function with $e=7$.
\item{FIG.~5}  The eigenenergy $E_{I=J}$ in GeV of the spin-isospin states
with $I=J=1/2$, 3/2 and 5/2 as a function of $L$  for various Skyrme term
constants $e$; (a) $e=3$, (b) $e=4$, (c) $e=5$ and  (d) $e=7$.
  For the conformal map profile functions, the energy of the symmetric state ,
anti-symmetric state  and almost degenerate state are represented by the
solid line, the dotted line and the solid line, respectively.
  The dot-dashed line represents the lowest eigenenergy for the arccosine
profile functions.
\item{FIG.~6} The $e$-dependence of the lowest eigenenergies $E_J$ for $J=0$,
 1/2, 3/2 and 5/2. The classical $S^3$ skyrmion energy $E_{cl}=6 \pi^2
f_{\pi}/e$
 and $E_{\infty}$ are also shown by dotted and dot-dashed lines.

\vfill\eject

\baselineskip=18 pt plus 1 pt minus 1 pt
\def\keih{\noalign{\hrule height1pt}}

\def\keill{\noalign{\hrule height0.1pt}}

\def\hf{\hfil\hskip 2mm}
\def\fh{\hskip 2mm \hfil}
\vskip 2 cm
{\offinterlineskip
\vskip 2mm
\def\sukid{&& && && && && && &&\cr}
          \centerline{TABLE I. \quad The lowest energies $E_J^{\rm min} $ in
MeV. \quad}

\vskip 3 mm
\halign{\strut#&\hf#\fh&#&\hf#\fh&#&\hf#\fh&#&\hf#\fh&#&\hf#\fh&#&\hf#\fh&#&\hf#\fh&#\cr
\keih
\sukid
& \quad $e$ \quad && $L$ \quad $E_0^{\rm min}$ && $L$ \quad $E_{1/2}^{\rm min}$
&& $L$ \quad $E_{3/2}^{\rm min}$ && $L$ \quad $E_{5/2}^{\rm min}$ && $E_{\rm
cl}$ && $E_{\infty}$ & \cr
\sukid
\keih
\sukid
& \quad 3 \quad && 1.1 \quad 1902 && 1.1 \quad  1931  && 1.2 \quad 2035  &&
1.4 \quad  2180 &&  1836 && 2596  & \cr
\sukid
\keill
\sukid
& \quad 4 \quad && 1.1 \quad 1456 && 1.2 \quad  1518  && 1.4  \quad 1712  &&
1.8  \quad  1928 &&  1377 && 1947  & \cr
\sukid
\keill
\sukid
& \quad 5 \quad && 1.2 \quad  1185  && 1.4 \quad  1294 && 1.8  \quad
$1566^{\dag}$  &&  2.5 \quad $1837^{\dag}$  && 1101  && 1558  & \cr
\sukid
\keill
\sukid
& \quad 7 \quad && 1.3 \quad  870  && 1.8  \quad  1090  && 2.7 \quad
$1455^{\dag}$  && 3.5 \quad $1750^{\dag}$  && \quad 787  &&  1113   & \cr

\sukid
\keih
}
}
\vskip 0.5 cm
\* The state a $\dag$ is attached to is not stable and will infinitely disperse
 when $L$ is varied.
\par
   \vfill\eject
\end